\documentclass[12pt]{article}

\begin{document}

\title{\textbf{The 3D Quantum Law of motion}}

\author{T.~Djama\thanks{E-mail:
{\tt djam\_touf@yahoo.fr}}}

\date {October 29, 2003}

\maketitle

\centerline {$14$, rue Si El Hou{\`e}s, B{\'e}ja{\"\i}a $06000$,
Algeria}

\begin{abstract}
\noindent In this paper, we give the solution of the three
dimensional quantum stationary Hamilton-Jacobi Equation
(3D-QSHJE) for a general form of the potential. We present the
quantum coordinates transformation with which the 3D-QSHJE takes
its classical form. Then, we derived the 3D quantum law of motion.
\end{abstract}

\vskip\baselineskip

\noindent PACS: 03.65.Bz; 03.65.Ca

\noindent Key words: quantum, Hamilton-Jacobi equation, three
dimensions, 3D, coordinates transformation.

\newpage

\vskip\baselineskip \noindent \textbf{1- Introduction}
\vskip\baselineskip

The deterministic interpretation of the quantum mechanics takes
none contesting progress during the last twenties years, and many
theoretical physicists contribute to this progress. Floyd took up
the quantum stationary Hamilton-Jacobi equation (QSHJE) \cite
{Flo1} already established by Bohm \cite {Bohm1} and de Broglie
\cite {Brog1}
\begin{eqnarray}
{1 \over 2m_0}\left({\vec{\nabla}S_0}\right)^2-{\hbar^2 \over
2m_0}{\Delta R \over R}+(E-V)=0\; \; , \hskip15mm (a)
\nonumber\\
\vec{\nabla} . \left(R^2\vec{\nabla}S_0\right)=0 \; \; .
\hskip35mm (b)
\end{eqnarray}
These equations arise from the 3D Schr{\"o}dinger equation, after
writing the wave function $\Psi$ in the form

\begin{equation}
\Psi(\vec{r})=R(\vec{r}) \left(\alpha e^{ {i \over \hbar}S_0}+
\beta e^{-{i \over \hbar}S_0}\right)\; .
\end{equation}
$R$ and $S_0$ are real functions. In the literature of
Hamilton-Jacobi formalism, $S_0$ represents the reduced action of
the studied system. Floyd focused his work on the one dimensional
(1D) motions and write the 1D QSHJE as \cite {Flo2}

\begin{eqnarray}
{1 \over 2m_0}\left({dS_0 \over dx}\right)^2- {\hbar^2 \over
4m_0}\left[{3 \over 2} \left({dS_0 \over dx}\right)^{-2}
\left({d^2 S_0 \over dx^2}\right)^2- \right. \hskip25mm\nonumber\\
\left. \left({dS_0 \over dx}\right)^{-1} \left({d^3 S_0 \over
dx^3}\right) \right]+ V(x)=E\; .
\end{eqnarray}
He established the solution of Eq. (3) in Ref. \cite {Flo2} and
used the Jacobi's theorem with its classical version in order to
plot the quantum trajectories  \cite {Flo1,Flo3}.

\noindent Recently, Djama and Bouda had criticized the use of the
classical version of Jacobi's theorem \cite {B-D1} and proposed a
new dynamical and deterministic approach to study the motions of
quantum particles \cite {B-D1,B-D2,B-D3,Flo4}. This new approach
consists on the introduction of a quantum Lagrangian from which we
deduce the fundamental expression of the quantum conjugate
momentum
\begin{equation}
\dot{x}{dS_0 \over dx}=2\; (E-V(x)) \; .
\end{equation}
We also derived the first integral of the quantum Newton's Law
(FIQNL) \cite {B-D1} and plotted the quantum trajectories \cite
{B-D2}. The Floyd quantum trajectories and ours are established
each either in one dimension, while the real motions are in 3D
spaces. So, a generalization in 3D of the deterministic approach
of quantum mechanics presented in Refs. \cite {B-D1,B-D2,B-D3} is
necessary.

The aim of the present paper is the resolution of the 3D-QSHJE
and deduce the 3D quantum law of motion. In this order, in Sec. 2,
we propose a solution of the 3D-QSHJE (Eqs. (1)). In Sec. 3, we
introduce a quantum coordinates transformation with which the
3D-QSHJE takes the form of the classical Hamilton-Jacobi equation
(CSHJE). Then, in Sec. 4, we derive the  quantum law of 3D
motion. Finally, in Sec. 5, we present a conclusion.

\vskip\baselineskip \noindent \textbf{2- The solution of the 3D
QSHJE } \vskip\baselineskip
 First, let us investigate Eq. (1.b). We can easily
check that it can be solved to give \cite {Bohm1,Flo5}

\begin{equation}
R^2(\vec r)\vec{\nabla}S_0=k\; (\phi \vec{\nabla}\theta\; -\theta
\vec{\nabla}\phi)\; ,
\end{equation}
where $k$ is a real constant. $\theta$ and $\phi$ are two real
and independent solutions of the 3D Schr{\"o}dinger equation
\begin{equation}
-{\hbar^2 \over 2m_0}\Delta\psi+\; V(\vec{r})=E\; .
\end{equation}
In fact, we can check that Eq. (5) is a consequence of the writing
the wave function in the form given by Eq. (2). In addition, Let
us write the reduced action $S_0$ in the form
\begin{equation}
S_0(\vec{r})=\hbar \arctan\left(a\ {\theta(\vec{r}) \over
\phi(\vec{r})}+b\right) \; .
\end{equation}
Replacing Eq. (7) into Eq. (5), we get
\begin{equation}
R(\vec{r})=\sqrt{{k \over \hbar a}}\; \;
\left[(a\theta+b\phi)^2+\phi^2\right]^{1 \over 2}\; .
\end{equation}
Expressions given by Eq. (7) and (8) are solutions of Eq. (1.b).
Are they solutions of Eq. (1.a)? In order to check this, we write
the reduced action $S_0$ in the form
\begin{equation}
S_0(\vec{r})=\hbar \arctan\left(\theta'(\vec{r}) \over
\phi(\vec{r})\right) \; ,
\end{equation}
where $\theta'(\vec{r})=a\; \theta(\vec{r})+b\; \phi(\vec{r})$.
Then, $R$ will be written as
\begin{equation}
R(\vec{r})=\sqrt{{k \over \hbar a}}\; \;
\left[\theta'^2+\phi^2\right]^{1 \over 2}\; .
\end{equation}
Now, Taking the usual derivative of $S_0$ and $R$ from Eqs. (9)
and (10), and replacing them into Eq. (1.a), we find
\begin{eqnarray}
{\hbar^2 \over 2m_0}{(\phi \vec{\nabla}\theta'\; -\theta'
\vec{\nabla}\phi)^2 \over (\theta'^2+\phi^2)^2}-{\hbar^2 \over
2m_0}{(\phi \vec{\nabla}\theta'\; -\theta' \vec{\nabla}\phi)^2
\over (\theta'^2+\phi^2)^2} \hskip20mm\nonumber\\
-{\hbar^2 \over2m_0}{\theta' \Delta \theta'+ \phi \Delta \phi
\over \theta'^2+\phi^2}+V(\vec{r})-E=0 \; ,
\end{eqnarray}
which reduces to
\begin{eqnarray}
{\theta'\over \theta'^2+\phi^2}\left[-{\; \hbar^2 \over 2m_0}\;
\Delta \theta'+(V(\vec{r})-E)\theta'\right]+ \hskip20mm\nonumber\\
{\phi \over \theta'^2+\phi^2}\left[-{\; \hbar^2 \over 2m_0}\;
\Delta \phi+(V(\vec{r})-E)\phi\right]=0\; .
\end{eqnarray}
Because it is a linear combination of $\theta$ and $\phi$,
$\theta'$ is a solution of the Schr{\"o}dinger equation (Eq. (6)).
This means that Eq. (12) is automatically satisfied. So,
expressions (7) and (8) of $S_0$ and $R$ are solutions of the
3D-QSHJE (Eqs. (1)).

\vskip\baselineskip \noindent \textbf{3- The quantum coordinates
and the classical form of the 3D-QSHJE} \vskip\baselineskip

In order to write the 1D-QSHJE with a similar form of the
classical Hamilton-Jacobi equation (CSHJE), Faraggi and Matone
have introduced the quantum coordinate ${\hat x}$ defined as
\cite {FM1}
\begin{equation}
\left({dx \over d{\hat x}}\right)^2=1-{\hbar^2 \over 2}\left({dS_0
\over dx }\right)^{-2}\; \{S_0,x\}
\end{equation}
After setting ${\hat S}_0=S_0$ and ${\hat V}=V$ they write the
QSHJE as
\begin{equation}
{1\over 2m_0}\left({d{\hat S}_0 \over d{\hat x}}\right)^2+{\hat
V}({\hat x})=E\; .
\end{equation}

That means that, in 3D, such a transformation, if it exist, will
permit to reduce the 3D-QSHJE in the form of the CSHJE. Before
going more, let us review some basic formula of the coordinates
transformation under a curved space.

\vskip\baselineskip \noindent \textbf{a- coordinates
transformation under curved space} \vskip\baselineskip

Let us consider a curved space with the spatial metric (here, we
do not consider time)
\medskip
\begin{equation}
ds^2=a_{\mu\nu}dx^{\mu}dx^{\nu}\; .
\end{equation}
$a_{\mu\nu}$ being the metric tensor, $x^{\mu}$ is the coordinate
with respect to the $\vec{e}_{\mu}$ direction. The well known
CSHJE can be written in such a space as \cite {Book}
\begin{equation}
a^{\mu\nu}({\bf x})(\partial_{\mu}S_0)(\partial_{\nu}S_0)+V({\bf
x})=E\; ,
\end{equation}
where $a^{\mu\nu}$ is the inverse tensor of $a_{\mu\nu}$, and
$\partial_{\mu}$ is the partial derivative with respect to
$x^\mu$. $S_0$ still represents the reduced action of the
particle.

\noindent Now, if we apply a coordinates transformation
$$
x^{\mu} \to x^{'\mu}\; ,
$$
the components of the metric tensor $a_{\mu\nu}$ and $a^{\mu\nu}$
will transform as
\begin{eqnarray}
\left\{ \begin{array}{cc}
               a_{\mu\nu}^{'}({\bf x^{'}})=\, {\partial x^{\alpha} \over \partial
x^{'\mu}}\, {\partial x^{\beta} \over \partial x^{'\nu}}\,
a_{\alpha\beta}({\bf x})\, ,\\ [.1in]
a^{'\mu\nu}({\bf x^{'}})=\,
{\partial x^{'\mu} \over \partial x^{\alpha} }\, {\partial
x^{'\nu} \over \partial x^{\beta} }\, a^{\alpha\beta}({\bf x})\, .
               \end{array}
       \right.
\end{eqnarray}
And the partial derivatives transform as
\begin{equation}
\partial_{\mu}^{'}=\, {\partial x^{\alpha} \over
\partial x^{'\mu}}\,  \partial_{\alpha}\, .
\end{equation}
Under this transformation, and after setting $S^{'}_0({\bf
x^{'}})=S_0({\bf x})$, the CSHJE (Eq. (16)) takes the form
\begin{equation}
a^{'\mu\nu}({\bf
x^{'}})(\partial^{'}_{\mu}S_0)(\partial^{'}_{\nu}S_0)+V({\bf
x({\bf x^{'}})})=E\; .
\end{equation}
In what follows, such transformation will be studied. It is the
transformation which reduce the 3D-QSHJE into its classical form.

\newpage
\noindent \textbf{b- quantum coordinates transformation}

Now, let us return to the 3D-QSHJE. We are looking for a
coordinate transformation
\begin{eqnarray}
\left\{ \begin{array}{cc}
               x \to {\hat x}(x,y,z)\\ [.1in]
               y \to {\hat y}(x,y,z)\\ [.1in]
               z \to {\hat z}(x,y,z)
               \end{array}
       \right.
\end{eqnarray}
by which and after setting
\[
\left\{ \begin{array}{cc}
               {\hat S}_0({\hat x},{\hat y},{\hat z})=S_0[\vec{r}({\hat x},{\hat
y},{\hat z}]\\ [.1in]
               {\hat V}({\hat x},{\hat y},{\hat z})=V[{\vec r}({\hat x},{\hat
y},{\hat z})]
               \end{array}
       \right.
\]
the 3D-QSHJE will be written in the classical form
\begin{equation}
{1 \over 2m_0}\left({\partial {\hat S}_0 \over
\partial {\hat x}}\right)^2+
{1 \over 2m_0}\left({\partial {\hat S}_0 \over
\partial {\hat y}}\right)^2+
{1 \over 2m_0}\left({\partial {\hat S}_0 \over
\partial {\hat z}}\right)^2+{\hat V}({\hat x},{\hat y},{\hat z})=E\;
.
\end{equation}
For this form, the corresponding metric tensor is the Euclidean
one
\begin{equation}
{\hat a}^{\mu\mu}=1\; , \hskip5mm {\hat a}^{\mu\nu}_{\mu \neq
\nu}=0 \hskip5mm \mu , \nu=1, 2 ,3 \; ,
\end{equation}
with $1$, $2$ and $3$ correspond to the directions of ${\hat x}$,
${\hat y}$ and ${\hat z}$. In what follows, we demonstrate the
existence of the transformation (20). So, using relations (18),
we write
\begin{eqnarray}
\left\{ \begin{array}{cc}
               {\partial \  \over \partial {\hat x}}={\partial \  \over
\partial  x}{\partial x \over \partial {\hat x}}+{\partial \  \over
\partial  y}{\partial y \over \partial {\hat x}}+{\partial \  \over
\partial  z}{\partial z \over \partial {\hat x}}\\ [.1in]
               {\partial \  \over \partial {\hat y}}={\partial \  \over
\partial  x}{\partial x \over \partial {\hat y}}+{\partial \  \over
\partial  y}{\partial y \over \partial {\hat y}}+{\partial \  \over
\partial  z}{\partial z \over \partial {\hat y}}\\ [.1in]
               {\partial \  \over \partial {\hat z}}={\partial \  \over
\partial  x}{\partial x \over \partial {\hat z}}+{\partial \  \over
\partial  y}{\partial y \over \partial {\hat z}}+{\partial \  \over
\partial  z}{\partial z \over \partial {\hat z}}\; .
               \end{array}
       \right.
\end{eqnarray}
\medskip
Replacing these last relations into Eq. (21), we find
\medskip
\begin{eqnarray}
{1 \over 2m_0}\left({\partial  S_0 \over
\partial  x}\right)^{2}\; a^{xx}\, +\; {1
\over 2m_0}\left({\partial  S_0 \over
\partial  y}\right)^{2}.\; a^{yy}
\, \, +{1 \over 2m_0}\left({\partial  S_0 \over
\partial  z}\right)^{2}\; a^{zz}\,+
\hskip4mm\nonumber\\
 {1 \over m_0}{\partial  S_0 \over
\partial  x}{\partial  S_0 \over
\partial  y}a^{xy}
+\; {1 \over m_0}{\partial  S_0 \over
\partial  x}{\partial  S_0 \over
\partial  z}a^{xz}
+{1 \over m_0}{\partial  S_0 \over
\partial  z}{\partial  S_0 \over
\partial  y}a^{yz}
+\, V(\vec{r})=E \; ,
\end{eqnarray}
where
\begin{eqnarray}
\left\{ \begin{array}{cc}
               a^{xx}=\left({\partial x \over \partial {\hat x}}\right)^{2}+
\left({\partial x \over \partial {\hat y}}\right)^{2}+
\left({\partial x \over \partial {\hat z}}\right)^{2}\\ [.1in]
               a^{yy}=\left({\partial y \over \partial {\hat x}}\right)^{2}
 + \left({\partial y \over \partial {\hat y}}\right)^{2}+
 \left({\partial y \over \partial {\hat z}}\right)^{2}\\ [.1in]
               a^{zz}=\left({\partial z \over \partial {\hat
x}}\right)^{2}+ \left({\partial z \over \partial {\hat
y}}\right)^{2}+ \left({\partial z \over \partial {\hat
z}}\right)^{2}\\ [.1in] a^{xy}=a^{yx}={\partial x \over
\partial {\hat x}} {\partial y \over
\partial {\hat x}}+ {\partial x \over \partial {\hat y}}
{\partial y \over \partial {\hat y}}+{\partial x \over
\partial {\hat z}} {\partial y \over \partial {\hat z}}\\ [.1in]
a^{xz}=a^{zx}={\partial x \over \partial {\hat x}}{\partial z
\over
\partial {\hat x}}+ {\partial x \over \partial {\hat y}} {\partial z \over
\partial {\hat y}}+{\partial x \over
\partial {\hat z}} {\partial z \over \partial {\hat z}}\\ [.1in]
a^{yz}=a^{zy}={\partial z \over \partial {\hat x}} {\partial y
\over\partial {\hat x}}+ {\partial z \over \partial {\hat y}}
{\partial y \over \partial {\hat y}}+{\partial z \over
\partial {\hat z}} {\partial y \over \partial {\hat z}} \; .
               \end{array}
       \right.
\end{eqnarray}
The quantities $a^{xy}$ are the components of the metric tensor of
the quantum curved space, where the 3D-QSHJE is described.

Eq. (24) represents the 3D-QSHJE written with the coordinates $x$,
$y$ and $z$, then it must be equivalent to Eq. (1.a). A
comparison of Eqs. (24) and (1.a) permit to define the quantities
$\partial x^{\mu}/\partial {\hat x}^{\nu}$. Indeed, if we write
Eq. (1.a) as
\begin{eqnarray}
{1 \over 2m_0} \left(\partial S_0 \over \partial
x\right)^2\left[1-\hbar^2\left(\partial S_0 \over \partial
x\right)^{-2}{\partial^2 R/ \partial x^2 \over R}\right]+
\hskip35mm\nonumber\\
{1 \over 2m_0} \left(\partial S_0 \over \partial y\right)^2
\left[1- \hbar^2\left(\partial S_0 \over \partial
y\right)^{-2}{\partial^2 R/ \partial y^2 \over R}\right]+
\hskip25mm\nonumber\\
{1 \over 2m_0} \left(\partial S_0 \over \partial
z\right)^2\left[1-\hbar^2\left(\partial S_0 \over
\partial z\right)^{-2}{\partial^2 R/ \partial z^2 \over
R}\right]+V(\vec{r})=E\;
\end{eqnarray}
and compare it with Eq.(26), we find
\begin{eqnarray}
\left\{ \begin{array}{cc}
               a^{xx}=\left({\partial x \over \partial {\hat x}}\right)^{2}+
\left({\partial x \over \partial {\hat y}}\right)^{2}+
\left({\partial x \over \partial {\hat
z}}\right)^{2}=1-\hbar^2\left(\partial S_0 \over \partial
x\right)^{-2}{\partial^2 R/ \partial x^2 \over R}\\ [.1in]
               a^{yy}=\left({\partial y \over \partial {\hat x}}\right)^{2}+
\left({\partial y \over \partial {\hat y}}\right)^{2}+
\left({\partial y \over \partial {\hat
z}}\right)^{2}=1-\hbar^2\left(\partial S_0 \over \partial
y\right)^{-2}{\partial^2 R/ \partial y^2 \over R}\\ [.1in]
                   a^{zz}=\left({\partial z \over \partial {\hat
x}}\right)^{2}+ \left({\partial z \over \partial {\hat
y}}\right)^{2}+ \left({\partial z \over \partial {\hat
z}}\right)^{2}=1-\hbar^2\left(\partial S_0 \over \partial
z\right)^{-2}{\partial^2 R/ \partial z^2 \over R} \\[.1in]
               a^{xy}=a^{yx}={\partial x \over
\partial {\hat x}} {\partial y \over
\partial {\hat x}}+ {\partial x \over \partial {\hat y}}
{\partial y \over \partial {\hat y}}+{\partial x \over
\partial {\hat z}} {\partial y \over \partial {\hat z}}=0\\ [.1in]
               a^{xz}=a^{zx}={\partial x \over \partial {\hat x}}{\partial z
\over
\partial {\hat x}}+ {\partial x \over \partial {\hat y}} {\partial z \over
\partial {\hat y}}+{\partial x \over
\partial {\hat z}} {\partial z \over \partial {\hat z}}=0\\[.1in]
            a^{yz}=a^{zy}={\partial z \over \partial {\hat x}} {\partial y
\over
\partial {\hat x}}+ {\partial z \over \partial {\hat y}}
{\partial y \over \partial {\hat y}}+{\partial z \over
\partial {\hat z}} {\partial y \over \partial {\hat z}}=0 \;
.
               \end{array}
       \right.
\end{eqnarray}
The set (27) of equations permit to define partially the quantum
transformation that we are searching for. Indeed, to define these
transformation, we must determine the nine quantities $\partial
x^{\mu}/\partial {\hat x}^{\nu}$. However, Eq. (25) contains six
equalities and nine unknown quantities $\partial x^{\mu}/\partial
{\hat x}^{\nu}$. Then, from Eq. (27), we can just determine six of
these unknown. In order to determine the three other quantities,
we use the tensor transformation relations given in Eq. (17) and
taking into account the relations (22), we write six equations
containing three additive unknown quantities which are the
components $a_{\mu\mu} (\mu=1,2,3)$ ($a_{\mu\nu}=0\;
(\mu\neq\nu)$)
\begin{eqnarray}
\left\{ \begin{array}{cc}
               \left({\partial x \over \partial {\hat x}}
\right)^{2}a_{xx}+\left({\partial y \over \partial {\hat x}}
\right)^{2}a_{yy}+\left({\partial x \over \partial {\hat x}}
\right)^{2}a_{zz}=1\; ,\\ [.1in]
               \left({\partial x \over \partial {\hat y}}
\right)^{2}a_{xx}+\left({\partial y \over \partial {\hat y}}
\right)^{2}a_{yy}+\left({\partial x \over \partial {\hat y}}
\right)^{2}a_{zz}=1\; ,\\ [.1in]
                   \left({\partial x \over \partial {\hat z}}
\right)^{2}a_{xx}+\left({\partial y \over \partial {\hat z}}
\right)^{2}a_{yy}+\left({\partial x \over \partial {\hat z}}
\right)^{2}a_{zz}=1\; , \\[.1in]
              {\partial x \over \partial {\hat x}}{\partial x \over \partial
{\hat y}}\, a_{xx}+{\partial y \over \partial {\hat x}}{\partial y
\over \partial {\hat y}}\, a_{yy}+{\partial z \over \partial {\hat
x}}{\partial z \over \partial {\hat y}}\, a_{zz}=0\; , \\ [.1in]
              {\partial x \over \partial {\hat x}}{\partial x \over \partial
{\hat z}}\, a_{xx}+{\partial y \over \partial {\hat x}}{\partial y
\over \partial {\hat z}}\, a_{yy}+{\partial z \over \partial {\hat
x}}{\partial z \over \partial {\hat z}}\, a_{zz}=0\; , \\[.1in]
        {\partial x \over \partial {\hat z}}{\partial x \over \partial
{\hat y}}\, a_{xx}+{\partial y \over \partial {\hat z}}{\partial y
\over \partial {\hat y}}\, a_{yy}+{\partial z \over \partial {\hat
z}}{\partial z \over \partial {\hat y}}\, a_{zz}=0\; .
               \end{array}
       \right.
\end{eqnarray}
Finally, we have twelve equations containing twelve unknown
quantities. Thus, the quantum transformation, that we searching
for really exist and is deduced from Eqs. (27) and (28). In
conclusion, we stress that this transformation reduces the
3D-QSHJE into its classical form (Eq. (21)).

\vskip\baselineskip \noindent \textbf{4- The quantum Lagrangian
and the quantum law of motion} \vskip0.5\baselineskip

The Use of the quantum coordinates bring the idea that the
quantum Lagrangian must have the same form as the classical one,
but written with the 3D quantum coordinates. So, we write
\begin{equation}
L_q={m_0 \over 2}\dot{\hat x}^2(x,y,z)+{m_0 \over 2}\dot{\hat
y}^2(x,y,z)+{m_0 \over 2}\dot{\hat z}^2(x,y,z)-{\hat V}({\hat
x},{\hat y},{\hat z})
\end{equation}
The total differentials $d{\hat x}$, $d{\hat y}$ and $d{\hat z}$
might be expressed with the partial derivatives $\partial {\hat
x}^{\mu} / \partial x^{\nu}$ and the total differentials $dx$,
$dy$ and $dz$ as
$$
d{\hat x}^{\mu}={\partial {\hat x}^{\mu} \over \partial
x^{\nu}}\, dx^{\nu}\hskip10mm  \mu, \, \nu =1,2,3.
$$
 Replacing the last relations into Eq. (29), we find
\begin{equation}
L_q(x^{\mu}, \dot{x}^{\mu})={m_0 \over 2}\dot
{x}^2a_{xx}(\vec{r})+{m_0 \over 2}\dot{y}^2a_{yy}(\vec{r})+{m_0
\over 2}\dot{z}^2a_{zz}(\vec{r})- V(\vec{r})\; ,
\end{equation}
where $a_{\mu\mu}$ are the diagonal components of the inverse of
the metric tensor ($a_{\mu\nu(\mu \neq \nu)}=0 $). They satisfy
the following relations
\begin{eqnarray}
\left\{ \begin{array}{cc}
               a_{xx}={1 \over a^{xx}}=\left({\partial {\hat x} \over \partial x}
\right)^{2}+\left({\partial {\hat y} \over \partial x}
\right)^{2}+\left({\partial {\hat z} \over \partial
x}\right)^{2}\\ [.1in]
               a_{yy}= {1 \over a^{yy}}=\left({\partial {\hat x} \over \partial
y} \right)^{2}+\left({\partial {\hat y} \over \partial y}
\right)^{2}+\left({\partial {\hat z} \over \partial
y}\right)^{2}\\ [.1in]
               a_{zz}= {1 \over a^{zz}}=\left({\partial {\hat x} \over \partial
z} \right)^{2}+\left({\partial {\hat y} \over \partial z}
\right)^{2}+\left({\partial {\hat z} \over
\partial z}\right)^{2}\; .
               \end{array}
       \right.
\end{eqnarray}
Expression (30) of the 3D quantum Lagrangian is analogous to the
one exhibited in Ref. \cite {B-D1} for the 1D cases where the
function $f(x,a,b)$ plays the same role as the components of the
metric tensor of the quantum curved space.

Now, using the expression (30) of the Lagrangian, the least action
principle leads to
\begin{eqnarray}
\left\{ \begin{array}{cc}
               m_0\ddot{x}a_{xx}+m_0\dot{x}{da_{xx} \over dt}-{m_0 \over
2}\dot{x}^2{\partial a_{xx} \over \partial x}-{m_0 \over
2}\dot{y}^2{\partial a_{yy} \over \partial x}-{m_0 \over
2}\dot{z}^2{\partial a_{zz} \over \partial x}+{\partial V \over
\partial x}=0\; ,\\ [.1in]
              m_0\ddot{y}a_{yy}+m_0\dot{y}{da_{yy} \over dt}-{m_0 \over
2}\dot{x}^2{\partial a_{xx} \over \partial y}-{m_0 \over
2}\dot{y}^2{\partial a_{yy} \over \partial y}-{m_0 \over
2}\dot{z}^2{\partial a_{zz} \over \partial y}+{\partial V \over
\partial y}=0\; ,\\ [.1in]
                m_0\ddot{z}a_{zz}+m_0\dot{z}{da_{zz} \over dt}-{m_0 \over
2}\dot{x}^2{\partial a_{xx} \over \partial z}-{m_0 \over
2}\dot{y}^2{\partial a_{xx} \over \partial z}-{m_0 \over
2}\dot{z}^2{\partial a_{xx} \over \partial z}+{\partial V \over
\partial z}=0\; .
               \end{array}
       \right.
\end{eqnarray}
Summing Eqs. (32), we get after calculation
\begin{eqnarray}
{m_0 \over 2}\; d(\dot{x}^2)\, a_{xx}+ {m_0 \over 2}\;
d(\dot{y}^2)\, a_{yy}+ {m_0 \over 2}\; d(\dot{z}^2)\, a_{zz}+
\hskip15mm\nonumber\\
{m_0 \over 2}\; \dot{x}^2\, da_{xx}+ {m_0 \over 2}\; \dot{y}^2\,
da_{yy}+ {m_0 \over 2}\; \dot{z}^2\, da_{zz}+dV=0\; ,
\end{eqnarray}
which, after integrating, reduces to
\begin{equation}
{m_0 \over 2}\; \left [\dot{x}^2\, a_{xx}+ \; \dot{y}^2\, a_{yy}+
\; \dot{z}^2\, a_{zz}\right ]+V(\vec{r})=E\; .
\end{equation}
$E$ being an integration constant representing the energy of the
particle. Eq. (34) represents the 3D quantum conservation
equation. Comparing Eqs. (26) and (34), and taking into account of
Eqs. (27), we deduce
\begin{eqnarray}
\left\{ \begin{array}{cc}
               a_{xx}={\partial S_0 / \partial x \over m_0\dot{x}}\; ,\\ [.1in]
               a_{yy}={\partial S_0 / \partial y \over m_0\dot{y}}\; ,\\ [.1in]
               a_{zz}={\partial S_0 / \partial z \over m_0\dot{z}}\; .
               \end{array}
       \right.
\end{eqnarray}
Replacing Eqs. (35) into Eq. (34), we find
\begin{equation}
{\dot x}{\partial S_0 \over \partial x}+{\dot y}{\partial S_0
\over \partial y}+{\dot z}{\partial S_0 \over \partial
z}=2\left[E-V(\vec{r})\right]\; .
\end{equation}
Eq. (36) expresses the relation between the quantum conjugate
momenta and the components of the speed of the particle. It can be
written, with vectors as
\begin{equation}
\vec{v}\; . \; \vec{\nabla S_0}=2\left[E-V(\vec{r})\right]\; ,
\end{equation}
\medskip where $ \vec{v}={\dot x}\vec{i}+{\dot y}\vec{j}+{\dot z}\vec{k}
$ is the 3D speed of the particle and
$$\vec{\nabla S_0}={\partial S_0 \over \partial x}\vec{i}+
{\partial S_0 \over \partial y}\vec{j}+{\partial S_0 \over
\partial z}\vec{k}$$
is the 3D quantum conjugate momentum.

\noindent From Eqs. (36) and (37), and, knowing the two
independent solutions of the Schr{\"o}dinger equation, one can
plot the quantum trajectories in 3D. So, Eqs. (36) and (37) might
be considered as a fundamental law of 3D quantum motions. The 3D
quantum Newton's Law can be derived from Eqs. (36) and (37),
after using the 3D-QSHJE, with the same manner as it is done in
Ref. \cite {B-D1} for the 1D quantum Newton's Law. That will be
investigated in a next paper. Remark that at the classical limit
($\hbar \to 0 $), $a_{\mu\mu}$ and $a^{\mu\mu}$ reduces to $1$
(see Eqs. (27)), then, from Eqs. (35), we deduce that
$$
{\partial S_0 \over \partial x}=m_0\dot{x}\hskip7mm {\partial S_0
\over
\partial y}=m_0\dot{y}\hskip5mm {\rm and} \hskip5mm{\partial S_0 \over
\partial z}=m_0\dot{z}\; .
$$
Taking these last relations into Eq. (36), we find the well known
classical conservation equation

$$
{m_0 \over 2}(\dot{x}^2+\dot{y}^2+\dot{z}^2)+V(\vec{r})=E
$$
Another important remark is that, when the particle have not
motions along $y$ and $z$ axis, Eqs. (36) and (37) reduce to the
expression (4) of the 1D conjugate momentum.

\vskip\baselineskip \noindent \textbf{5- Conclusion}
\vskip0.25\baselineskip

The resolution of the quantum stationary Hamilton-Jacobi equation
in three dimensions (Eqs. (1)) is a very important realization in
the frame of the construction of a deterministic approach of
quantum mechanics. The reduced action $S_0(\vec{r})$ that we
proposed as solution of the 3D-QSHJE has the same form as the one
dimensional reduced action already established by Floyd \cite
{Flo2} and Faraggi and Matone \cite {FM1}.

\noindent The introduction of a quantum coordinates
transformation reducing the 3D-QSHJE into its classical form
constitute a hopeful step to construct a dynamical approach of
quantum mechanics. This transformation give us the idea about the
manner with which the dynamical approach of quantum motions in
three dimension must be constructed. Indeed, in 1D, we have
constructed in Ref. \cite {B-D1} a quantum Lagrangian and
Hamiltonian from which we established the expression of the
quantum conjugate momentum, and derived the Quantum Newton's Law.
For 3D cases, a similar construction is possible now. That is
what we investigate in this paper.

\noindent The quantum law of 3D motions that we exhibited in Eqs.
(36) and (37) is the generalization to the 3D spaces of the
quantum Law of motion that we have already established with Bouda
in Ref. \cite {B-D1}. For this deterministic and causal approach
of quantum mechanics such a generalization represents an
important advancement of theory, since all possible physical
phenomena happen in 3D spaces. The fundamental relation expressed
in Eqs. (36) and (37) contains, in the left hand side, the scalar
product of the conjugate momentum and the velocity, and in the
right hand side, the kinetic energy of the particle up a
multiplicative constant.

Finally, we stress that the generalization into 3D problems of
the deterministic approach, that we have already introduced in
Ref. \cite {B-D1}, can be seen as the first step to the
investigation of the quantum gravity in the framework of a causal
quantum theory.

\vskip\baselineskip
\noindent \textbf{REFERENCES} 

\end{document}